\documentclass[floatfix,
amsmath,amssymb,aps,twocolumn,superscriptaddress,prl]{revtex4}

\usepackage{amsfonts}
\usepackage[pdftex]{graphicx}
\usepackage{color}
\usepackage{bm}
\usepackage{multirow}

\newcommand{\vect}[1]{{\mathbf #1}}
  
\newcommand{\up}{\uparrow}
\newcommand{\down}{\downarrow}
\renewcommand{\k}{{\bf k}}

\newcommand{\q}{{\bf q}}

\newcommand{\eb}{\epsilon_b}

\newcommand{\nn}{\nonumber}

\newcommand{\T}{{\cal T}}
\newcommand{\F}{{\cal F}}
\newcommand{\Ham}{{\cal H}}
\newcommand{\op}{{\omega_z}}
\begin{document}


\title{Bound states in a quasi-two-dimensional Fermi gas}

\author{Jesper Levinsen}
\affiliation{T.C.M. Group, 
Cavendish Laboratory, JJ Thomson Avenue, Cambridge,
 CB3 0HE, United Kingdom} %

\author{Meera M. Parish}
\affiliation{London Centre for Nanotechnology, Gordon Street, London, WC1H 0AH, United Kingdom}

\date{\today}

\begin{abstract}
We consider the problem of $N$ identical fermions of mass
$m_\uparrow$ and one distinguishable particle of mass $m_\downarrow$
interacting via short-range interactions in a confined
quasi-two-dimensional (quasi-2D) geometry.  For $N=2$ and mass
ratios $m_\uparrow/m_\downarrow <13.6$, we find non-Efimov trimers
that smoothly evolve from 2D to 3D. In the limit of strong 2D
confinement, we show that the energy of the $N+1$ system can be
approximated by an effective two-channel model.  We use this
approximation to solve the $3+1$ problem and we find that a bound
tetramer can exist for mass ratios $m_\uparrow/m_\downarrow$ as low
as 5 for strong confinement, thus providing the first example of a
universal, non-Efimov tetramer involving three identical fermions.
\end{abstract}


\maketitle

An understanding of the few-body problem can be important for gaining
insight into the many-body system.  In dimensions higher than one,
few-body bound states can, for instance, impact the statistics of the
many-body quasiparticle excitations. Indeed, for fermionic systems,
the two-body bound state is fundamental to the understanding of the
BCS-BEC crossover~\cite{Leggett1980,Comte,nozieres1985,sademelo1993},
while the existence of three-body bound states of
fermions~\cite{trimer,PhysRevLett.103.153202} with unequal masses can
lead to dressed trimer quasiparticles in the highly polarized Fermi
gas~\cite{Mathy:2011ys}.  Even in one dimension (1D), few-body bound
states can impact the many-body phase: It has already been shown that
one can have a Luttinger liquid of trimers~\cite{orso2010}.

In general, attractively interacting bosons readily form bound
clusters, with the celebrated example being the Efimov effect in 3D
\cite{Efimov}. Here, there is a universal hierarchy of trimer states
for resonant short-range interactions, while clusters of four or more
bosons can also form
\cite{Hammer2007,stecher2009,Ferlaino2008,stecher2010}. Even in the
limit of a 2D geometry, where the Efimov effect is absent, both
trimers \cite{Bruch1979} and tetramers \cite{Platter2004} have been
predicted. On the other hand, bound states of identical fermions are
constrained to have odd angular momentum owing to Pauli exclusion and
thus, even for attractive interactions, identical fermions are subject
to a centrifugal barrier. For short-range $s$-wave~\footnote{Trimers
  containing three identical fermions interacting {\em via}
  short-range $p$-wave interactions have also been predicted, see
  M. Jona-Lasinio, L. Pricoupenko, and Y. Castin, Phys. Rev. A 77,
  043611 (2008).}  interactions in 3D, non-Efimov trimers consisting
of two identical fermions with mass $m_\up$ and one distinguishable
particle with mass $m_\down$ can only exist above the critical mass
ratio $m_\up/m_\down \simeq 8.2$~\cite{trimer}, while Efimov trimers
only appear once $m_\up/m_\down \gtrsim 13.6$~\cite{LL}. However, the
existence of larger $(N+1)$-body bound states involving $N>2$
identical fermions remains largely unknown --- it has only recently
been shown that Efimov tetramers exist in
3D~\cite{PhysRevLett.105.223201}.

\begin{figure}
\centering
\includegraphics[width=0.9\linewidth]{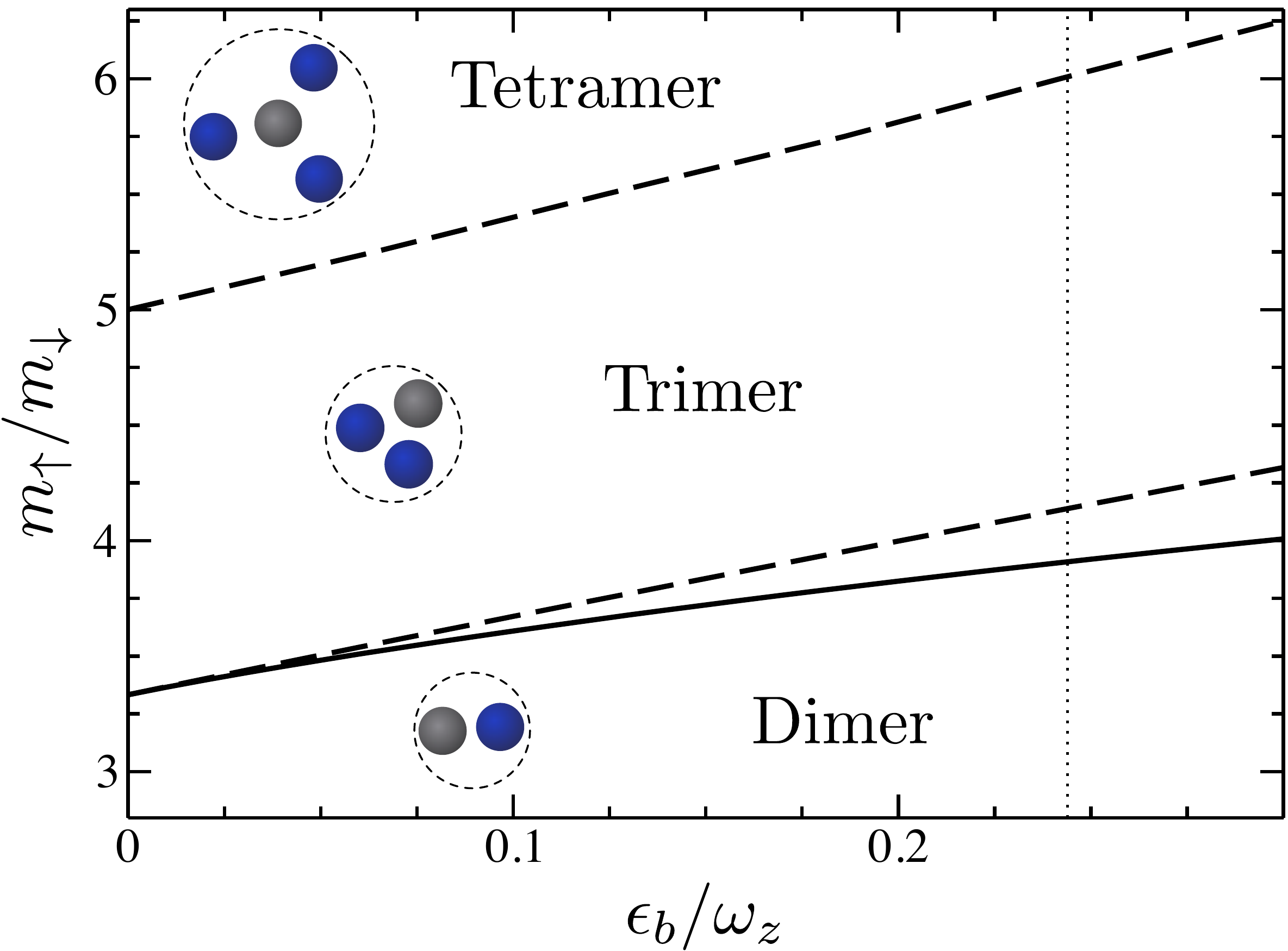}
\caption{(Color online) Critical mass ratio for the appearance of
  trimers and tetramers in quasi-2D, where the 2D limit corresponds to
  $\epsilon_b/\omega_z \to 0$. The solid line follows from the
  solution of the full three-body quasi-2D problem,
  Eq.~(\ref{eq:trimerEq}). Dashed lines follow from an effective
  two-channel model. The vertical dotted line marks unitarity, where
  the 3D scattering length diverges.}
\label{fig:phases}
\end{figure}

In this Letter, we investigate the problem of $N$ identical fermions
interacting with one distinguishable particle in a confined quasi-2D
geometry, where the centrifugal barrier is reduced and the binding of
fermions should be favored. Such 2D geometries have recently been
realised in ultracold atomic Fermi
gases~\cite{2DFermi_expt,PhysRevLett.106.105301,sommer2011_2D,Dyke2011,Zhang:2012uq},
where the fermions are confined to 2D with an effective harmonic
potential.  In addition to allowing one to explore the 2D-3D
crossover, the harmonic confinement can strongly modify the scattering
properties of atoms via confinement-induced
resonances~\cite{olshanii1998,PhysRevLett.103.153202,nishida2010}. It
has already been demonstrated that stable non-Efimov trimers can exist
for lower mass ratios $m_\up/m_\down$ in
quasi-2D~\cite{2Dtrimer,PhysRevLett.103.153202}.  Here we show that
tetramers involving $N=3$ identical fermions can appear for
$m_\up/m_\down$ as low as 5 in quasi-2D (see Fig.~\ref{fig:phases}),
thus putting it within reach of current cold-atom experiments.

We construct the general equations for the bound state of the $N+1$
system in quasi-2D and we reveal how to simplify the problem in the
case of the trimer ($N=2$).  In the limit of strong 2D confinement, we
show that the $N+1$ problem can be described by an effective
two-channel model, analogous to that used for Feshbach resonances.
This important simplification allows us to solve the aforementioned
$N=3$ problem in quasi-2D.

In the following, we assume the two atomic species $\{\up,\down\}$ to
be confined to a quasi-2D geometry by an approximately harmonic
potential along the $z$ direction,
$V_{\up,\down}(z)=\frac12m_{\up,\down}\omega_z^2z^2$.  Here, we
restrict ourselves to equal confinement frequencies for the two
species since it allows a separation of the relative and center of
mass motion along the $z$-direction, as we discuss below.  Such a
scenario can, in principle, be engineered experimentally using
spin-dependent optical lattices.  However, even in the case where the
confinement frequency is species dependent, regimes exist in which the
few-body properties are only weakly affected by this dependence. For
instance, for large mass ratios and on the molecular side of the
Feshbach resonance, once the $\up\down$ dimer is smaller than the
light atom oscillator length, $l_z^\down=\sqrt{\hbar/m_\down\op}$, the
light atom is essentially confined by its interaction with the heavy
atoms~\cite{PhysRevLett.103.153202}.

The starting point of our analysis is the $T$-matrix describing the
repeated two-body interspecies interaction. In the ultracold gases,
the interaction is described by a zero-range model as the van der
Waals range of the interatomic potential is much smaller than all
other length scales in the problem, including the confinement
lengths. The $T$-matrix may be considered in the basis of the
individual motion of a spin-$\down$ and $\up$ atom.  However, due to
the restriction to equal confinement frequencies for the two species,
the center of mass and relative motion separate and it is
advantageous to work in this basis.  In the center of mass frame of
the harmonic oscillator potential, at energy $\epsilon$ below the
two-body threshold $\op$ (we set $\hbar=1$) and at total 2D momentum
$\q$, the $T$-matrix takes the form \cite{petrov2001}
\begin{equation}
\T(\q,\epsilon)=\frac{\sqrt{2\pi}}{m_r}\left\{\frac{l_z^r}{a_s}-\F
\left(\frac{-\epsilon+\q^2/2(m_\up+m_\down)}{\op}\right)\right\}^{-1},
\label{eq:T}
\end{equation}
where the zero-range interaction is renormalized by the use of the 3D
scattering length, $a_s$. Here, $m_r=m_\up m_\down/(m_\up+m_\down)$ is
the reduced mass and $l_z^r=\sqrt{1/2m_r\omega_z}$ is the confinement
length corresponding to the relative motion.  We use the definition of
$\F$ \cite{RevModPhys.80.885}
\begin{equation}
\F(x)=\int_0^\infty\frac{du}{\sqrt{4\pi u^3}}\left(1-\frac{e^{-xu}}
{\sqrt{[1-\exp(-2u)]/2u}}\right).
\end{equation}
The two-dimensional scattering always admits a two-body bound state of
mass $M=m_\up+m_\down$ and binding energy $\eb>0$ satisfying
$l_z^r/a_s=\F(\eb/\omega_z)$.
 
The $T$-matrix in the basis of individual motion is related to $\T$
by the change of basis
\begin{eqnarray}
  T^{n_0'n_1'}_{n_0n_1}(\q,\epsilon) & = &
\sum_{n\,n_rn_r'\!\!}C^{n_0n_1}_{nn_r}(m_\down,m_\up)
C^{n_0'n_1'}_{nn_r'}(m_\down,m_\up) \nn \\ && \times
  \psi_{n_r}(0)\psi_{n_r'}(0)\T(\q,\epsilon-n\op).
\end{eqnarray}
Here, $n_0$ and $n_1$ are the quantum numbers labelling the
eigenstates of the single-particle Hamiltonians
$\Ham_{\down,\up}=-\frac{{\mathbf{\nabla}}_{0,1}^2}{2m_{\down,\up}}
+\frac12m_{\down,\up}\omega_z^2z_{0,1}^2$ while $n_r$ and $n$ are the
quantum numbers in the basis of relative, $z_{01}=z_0-z_1$, and center
of mass, $Z_{01}=(m_\down z_0+m_\up z_1)/M$, 
coordinates. The wavefunction of the relative motion takes the value
$\psi_{n_r}(0)=(-1)^{n_r/2}\sqrt{(n_r-1)!!/n_r!!}$ if $n_r$ is even,
and 0 otherwise. The Clebsch-Gordan coefficients
$C^{n_0n_1}_{nn_r}(m_\down,m_\up)\equiv\langle n_0n_1|nn_r\rangle$
were obtained in Ref.~\cite{Smirnov1962} and vanish unless
$n_0+n_1=n+n_r$.

\begin{figure}
\begin{center}
\includegraphics[width=0.9\linewidth,angle=0]{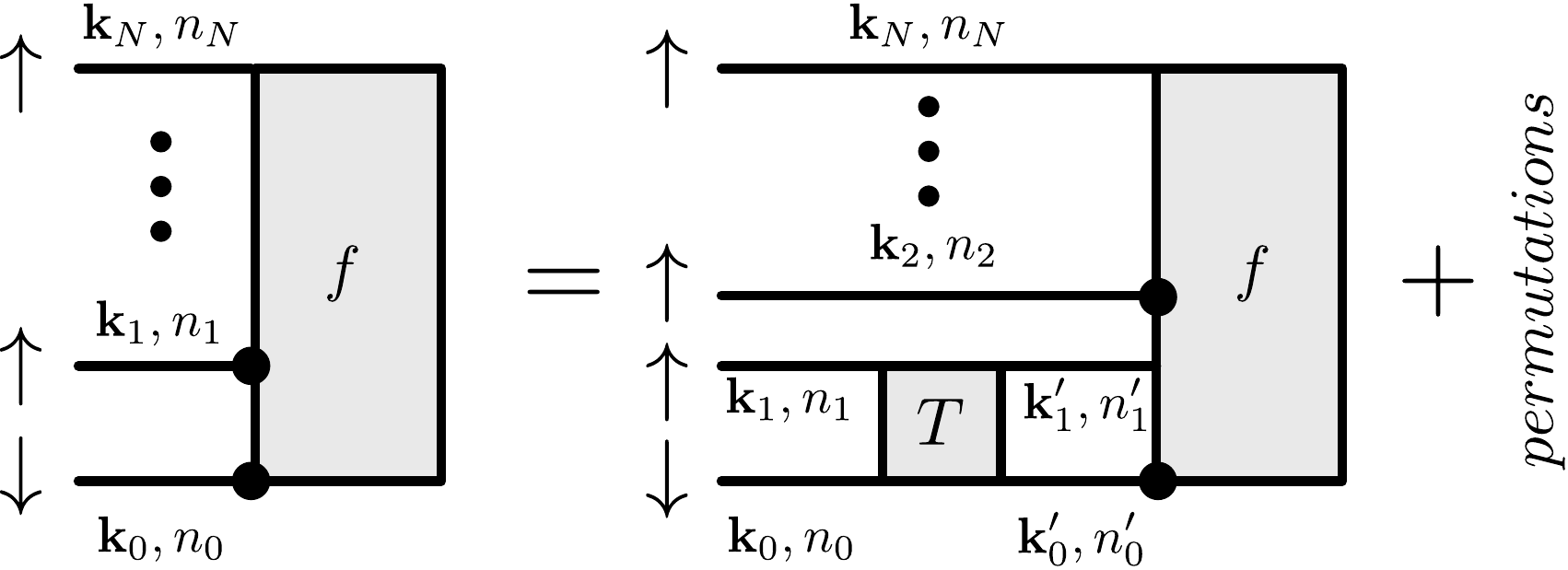}
\end{center}
\caption{The diagrams which give the binding energy of the $N+1$
  bound state in quasi-2D. Black dots indicate the initial interaction inside $f$.
}
\label{fig:multi0}
\end{figure}

We now turn to the question of the existence of bound states
consisting of $N$ spin-$\up$ atoms and a single spin-$\down$ atom. To
this end, we construct the sum of connected diagrams with $N+1$
incoming atoms (Fig.~\ref{fig:multi0}).  The $\up$ atoms are
considered on-shell with 2D momenta $\k_i$, harmonic oscillator
quantum numbers $n_i$, and corresponding single-particle energies
$\epsilon_{\k_in_i\up}=k_i^2/2m_\up+n_i\op$ for $i=1,\ldots,N$. We
consider scattering in the centre of mass frame of the 2D motion and
at a total energy $E$ below the $N+1$ atom threshold
$(N+1)\op/2$. Thus, the $\down$ atom has 2D momentum
$\k_0\equiv-\sum_{i=1}^N\k_i$, harmonic oscillator quantum number
$n_0$, and energy $E_0\equiv E-\sum_{i=1}^N \epsilon_{\k_in_i\up}$.
The sum of diagrams with $N+1$ incoming particles in which the $\down$
atom interacts first with the $\up$ atom numbered 1 is denoted
$f_{\k_2\ldots\k_N}^{n_0\ldots n_N}$. Note that there is no dependence
on $\k_1$ as the initial interaction depends only on the total
momentum of the two atoms.

The occurrence of a bound state corresponds to a singularity of $f$ at
its binding energy.  This singularity results from the summation of an
infinite number of diagrams and, at the pole, $f$ satisfies the
homogeneous integral equation illustrated in Fig.~\ref{fig:multi0}:
The initial interaction is described by a $T$-matrix, and then the
spin-$\down$ atom subsequently interacts with another of the $\up$
atoms. Thus, the right hand side contains $N-1$ terms and the integral
equation satisfied by the bound state energy is (setting the volume to
1):
\begin{eqnarray}
f_{\k_2\ldots\k_N}^{n_0\ldots n_N}
& = & 
-\sum_{\k_1',n_0'n_1'}\frac{T_{n_0n_1}^{n_0'n_1'}(\k_0+\k_1,
E_0+\epsilon_{\k_1n_1\up})}
{E_0+\epsilon_{\k_1n_1\up}-\epsilon_{\k_0'n_0'\down}-\epsilon_{\k_1'n_1'\up}}
\nn \\ &&\hspace{-12mm}
\times\left\{f^{n_0'n_2n_1'n_3\ldots
    n_N}_{\k_1'\k_3\ldots\k_N}+
\ldots+f^{n_0'n_Nn_2\ldots n_{N-1}n_1'}_{\k_2\ldots\k_{N-1}\k_1'}\right\},
\label{eq:np1}
\end{eqnarray}
where $\k_0'=\k_0+\k_1-\k_1'$, and the minus sign on the \emph{r.h.s.}
appears because $f$ is antisymmetric under the exchange of incoming
fermions.  Equation~\eqref{eq:np1} embodies a simple and generic
formulation for the $(N+1)$-body problem in quasi-2D, which in
principle allows us to capture the crossover from 2D to 3D.  Indeed,
for the case of $N=2$, it is a generalization of the
Skorniakov-Ter-Martirosian equation for atom-dimer
scattering~\cite{Skorniakov1956}, while for $N=1$, Eq.~\eqref{eq:np1}
simply reduces to the condition for the two-body binding
energy. Finally, we note that Ref.~\cite{Minlos} derived an expression
similar to our Eq.~(\ref{eq:np1}) for the 3D $N+1$ problem.

An important simplification to Eq.~\eqref{eq:np1} becomes possible in
the limit of strong quasi-2D confinement, $\op\gg \eb$.  Here, the
function $\F$ can be expanded as
\begin{equation}
  \F(x)\approx\frac1{\sqrt{2\pi}}\ln\left(\pi x/B \right)+\frac{\ln2}
{\sqrt{2\pi}}x+{\cal O}(x^2)
\label{eq:Gtaylor}
\end{equation}
with $B\approx0.905$ \cite{petrov2001,RevModPhys.80.885}.  On the
other hand, consider the denominator on the {\em r.h.s.} of
Eq.~(\ref{eq:np1}) which we shall write for simplicity as
$\epsilon-n\op$. Here, the typical energy scale $\epsilon\sim\eb$
since, for bound states, the function $f$ is strongly peaked at
momenta $\sim\sqrt{2m_r\eb}$, while it quickly decays for large
momenta.  Now, if we expand the denominator in powers of $\eb/\op$
(assuming $n\neq 0$), then the lowest order term vanishes when
integrated over momentum due to the antisymmetry of
$f_{\k_1'\ldots}$. Consequently, the lowest non-vanishing contribution
from the denominator is of order $(\eb/\op)^2$ when the harmonic
oscillator index $n$ is non-zero.  We conclude that to linear order in
$\eb/\op$ the integral equation for the $N+1$ bound state reduces to
\begin{eqnarray}
  f_{\k_2\ldots\k_N} & = &
  \tilde{\T}(\k_0+\k_1,\tilde E_0+\epsilon_{\k_1\up})
\nn \\ && \times
\sum_{\k_1'}\frac{f_{\k_1'\k_3\ldots\k_N}+
 \ldots+f_{\k_2\ldots\k_{N-1}\k_1'}}
{\tilde E_0+\epsilon_{\k_1\up}-\epsilon_{\k_0'\down}-\epsilon_{\k_1'\up}},
\label{eq:2channel}
\end{eqnarray}
with the single particle energies $\epsilon_{\k}\equiv
\epsilon_{\k0}$, $\tilde{E}_0=E-\sum_{i=1}^N\epsilon_{\k_i\up}$, and
$\tilde{\T}$ obtained from Eq.~(\ref{eq:T}) using the linear expansion
of $\F$, Eq.~(\ref{eq:Gtaylor}).  As the effects of confinement in
this limit are contained solely within the linearized $T$-matrix,
Eq.~(\ref{eq:2channel}) may be obtained through a strictly 2D
2-channel model~\cite{Timmermanns1999}, where the closed channel
corresponds to excited harmonic oscillator modes.  Thus, the
confinement length $l^r_z$ plays the role of an effective range in
this model, with the 2D limit $l^r_z/a_s \to 0$ corresponding to a
single-channel model.  This simplification crucially depends on the
antisymmetry resulting from Fermi statistics and it thus does not
apply to bound clusters involving bosons confined to 2D, as considered
in Refs.~\cite{brodsky2006,pricoupenko2011}.  Finally, we note that a
similar simplification was recently obtained for quasi-1D atom-dimer
scattering~\cite{Petrov2012} (see also Ref. \cite{mora2005}).

We now proceed to solve the three-body problem using the above
methods. First, note that using the zero-range condition and removing
the center of mass generally allows one to reduce the number of
harmonic oscillator quantum numbers by two in Eq.~(\ref{eq:np1})
\cite{PhysRevLett.103.153202}. For the three-body problem, this is
achieved by changing coordinates to the relative motion of the two
atoms initially interacting, $z_{01}$, 
the relative motion of
the pair and the third atom, $z_2^{01}=(m_\down z_0+m_\up
z_1)/(m_\down+m_\up)-z_2$, and the center of mass $Z_{012}=(m_\down
z_0+m_\up z_1+m_\up z_2)/(m_\down+2m_\up)$. Defining the corresponding
quantum numbers $n_{01}$, $n_2^{01}$, and $N_{012}$, we adopt the new
basis
$\chi_{\k_2}^{n_2^{01}}=\frac1{\psi_{n_{01}}(0)}\sum_{n_0n_1n_2}\langle
N_{012}n_2^{01}n_{01}|n_0n_1n_2\rangle f_{\k_2}^{n_0n_1n_2}
$~
\footnote{Here, we have dropped $n_{01}$ from the \emph{l.h.s.}
  as the resulting equation will be independent of this index. Also,
  the center of mass quantum quantum number, $N_{012}$, has been
  omitted as it only causes a shift in the energy; since we consider
  the lowest lying trimer, $N_{012}$ will be set to 0 in the
  following.}.
Then, Eq.~(\ref{eq:np1}) for the trimer becomes
\begin{eqnarray} \label{eq:trimerEq}
&&\hspace{-7mm}
\chi_{\k_2}^{n_2^{01} }= \T\left(\k_2,E-\epsilon_{\k_2\uparrow}
-n_2^{01}\op\right)
\nn \\ && \hspace{-6mm}\times\hspace{-4mm} \sum_{\k_1',n_1^{02}n_{02}n_{01}'}
\hspace{-5mm}
\frac{\psi_{n_{02}}(0)\psi_{n_{01}'}(0) \langle
  n_2^{01}n_{01}'|n_1^{02}n_{02}\rangle
\chi_{\k_1'}^{n_1^{02}}}
{E-\epsilon_{\k_1'\up}-\epsilon_{\k_2\up}-\epsilon_{\k_1'+\k_2\down}-
(n_1^{02}+n_{02})\op}. 
\end{eqnarray}
%
The matrix element in Eq.~\eqref{eq:trimerEq} may be
evaluated by a series of coordinate transformations:
\begin{eqnarray} 
&& \hspace{-4mm} \langle{n^{01}_2n_{01}}| n^{02}_1 n_{02} \rangle =
\sum_{n_0n_1 n_2N_{01}N_{02}} C^{N_{01} n_2}_{0
    n^{01}_2}(M
,m_\up)
\nn \\ \notag
&& \hspace{-3mm}\times \ C^{n_0 n_1}_{N_{01}
    n_{01}}(m_\down,m_\up) \ C^{n_0 n_2}_{N_{02}
    n_{02}}(m_\down,m_\up) 
C^{N_{02} n_1}_{0
    n^{02}_1}(M
,m_\up) \ ,
\end{eqnarray}
where several sums can be dropped due to the constraints on the
Clebsch-Gordan coefficients.

\begin{figure}
\centering
\includegraphics[width=0.85\linewidth]{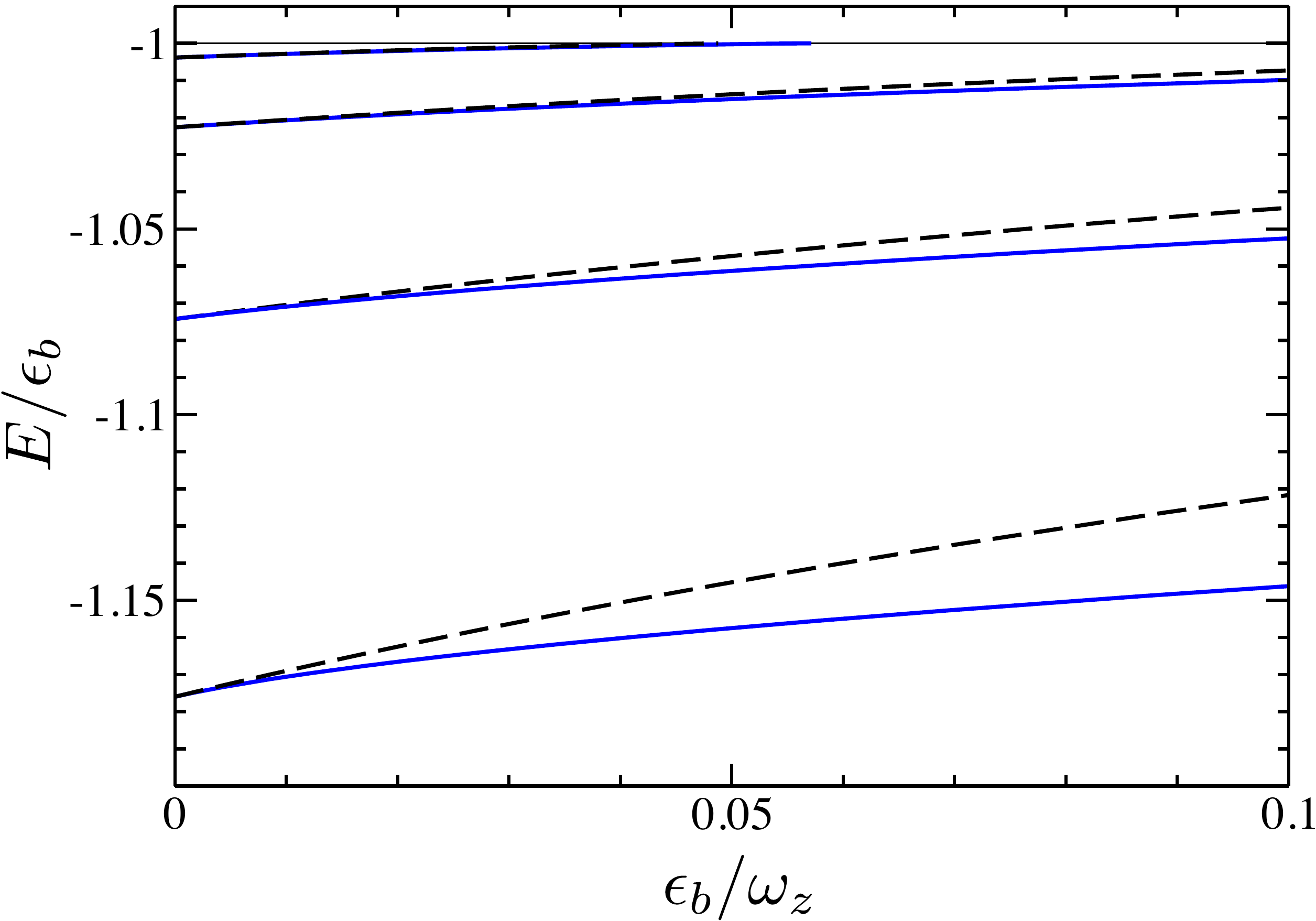}
\caption{(Color online) Energy of the trimer in quasi-2D for mass
  ratios $m_\up/m_\down =$ 3.5, 4, 5, 6.64 (from top to bottom).  The
  solid lines correspond to the full calculation, while the dashed
  lines are derived from the effective two-channel model,
  Eq.~(\ref{eq:2channel}).}
\label{fig:trimer}
\end{figure}

\begin{figure}
\centering
\includegraphics[width=0.85\linewidth]{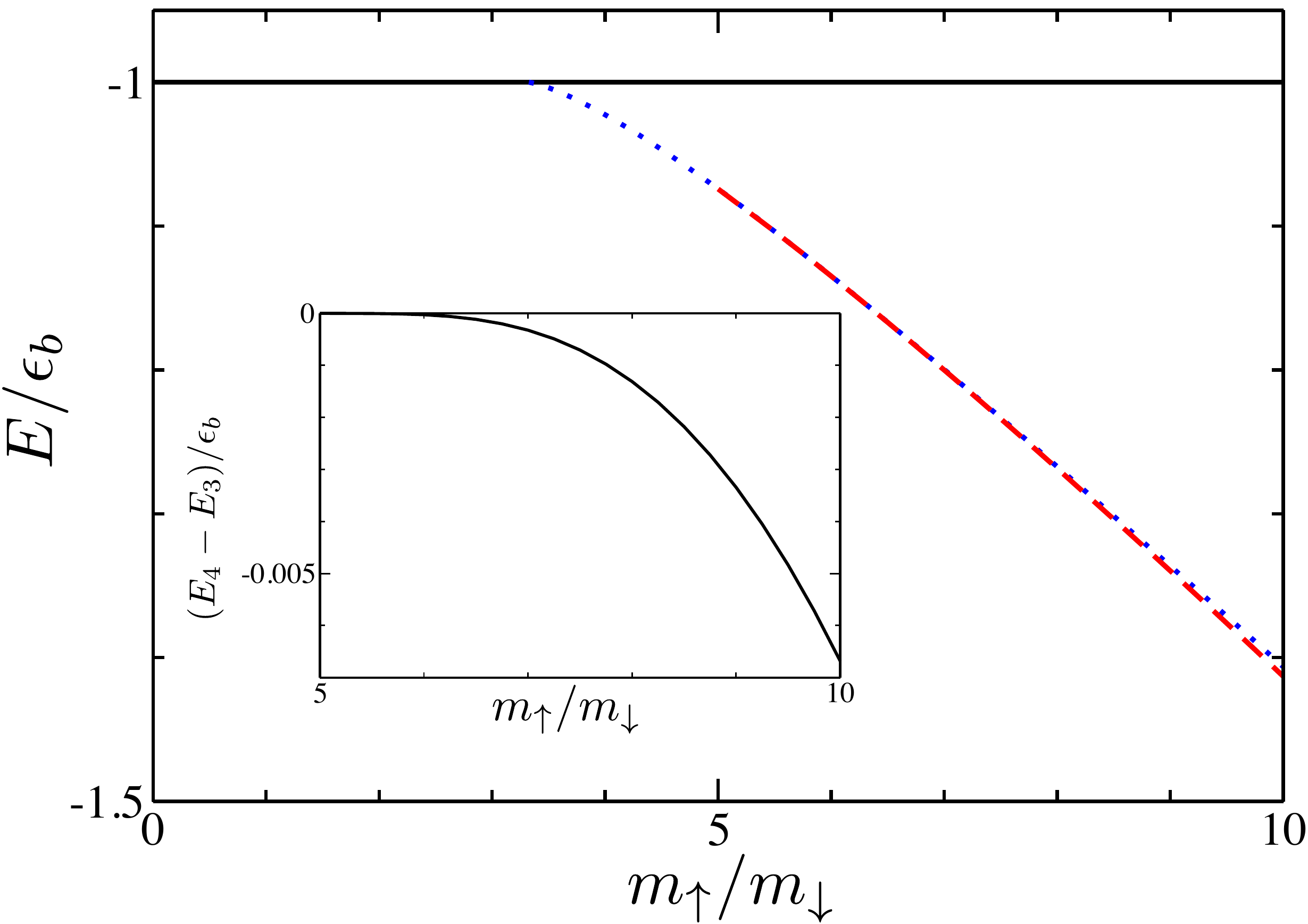}
\caption{(Color online) Energy of the dimer (solid line), trimer
  (dotted), and tetramer (dashed) in 2D as a function of mass
  ratio. The trimer binds when $m_\up/m_\down > 3.33$, consistent with
  Ref.~\cite{2Dtrimer}, while the trimer-tetramer transition occurs at
  $m_\up/m_\down = 5.0$.  {\em Inset:} The difference between trimer
  and tetramer energies, $E_3-E_4$.}
\label{fig:undressed}
\end{figure}

Since the trimer consists of identical fermions, it must necessarily
have odd angular momentum $L$ in the $x$-$y$ plane of the 2D
layer. Thus, the lowest-energy trimer has $L=1$, and it can be
regarded as a $p$-wave pairing of $\up$ fermions mediated by their
$s$-wave interactions with the light $\down$ particle.  In this case,
we have $\chi^{n^{01}_2}_{\vect{k}_2} = \tilde \chi^{n^{01}_2}_{k_2}
e^{i\phi_2}$, where $\phi_2$ is the angle of $\vect{k}_2$ with respect
to the $x$-axis and $\tilde \chi$ is a function of $k_2\equiv
|\vect{k}_2|$.  Integrating over $\phi_2$ in Eq.~\eqref{eq:trimerEq}
then leaves an integral equation that only depends on $k_2$ and
$n^{01}_2$.  The same applies for the two-channel model
Eq.~\eqref{eq:2channel} with $N=2$, where now there is only a
dependence on $k_2$.

We have calculated the trimer binding energy as a function of
confinement for a range of mass ratios, as depicted in
Fig.~\ref{fig:trimer}. We see that the binding energy decreases as we
perturb away from 2D and the centrifugal barrier is increased.
Correspondingly, we find that the critical mass ratio $m_\up/m_\down$
for the trimer binding increases as we perturb away from the 2D limit,
as shown in Fig. \ref{fig:phases}, and smoothly evolves towards the 3D
result of 8.2~\cite{trimer}. For the special case of $^{6}$Li-$^{40}$K
mixtures, where $m_\up/m_\down=6.64$, we have checked that our results
agree with Ref.~\cite{PhysRevLett.103.153202}.  In the limit of strong
2D confinement, we see that the two-channel model captures the
lowest-order dependence on $\epsilon_b/\omega_z$ of the trimer energy
and critical mass ratio.

We can exploit the two-channel model \eqref{eq:2channel} to solve the
more complicated four-body ($N=3$) problem in quasi-2D.  Once again,
the presence of identical fermions requires us to consider total
angular momentum $L=1$.  Thus, we have for the tetramer
\begin{align} \notag
 f_{\vect{k_2}\vect{k_3}}  
= \tilde f(k_2,k_3,\Delta\phi_{32}) e^{i\phi_2} = -\tilde
 f(k_3,k_2,-\Delta\phi_{32}) e^{i\phi_3},
\end{align}
where $\Delta\phi_{32} = \phi_3 - \phi_2$. We note that a similar
equation for the tetramer energy was obtained for the 3D problem in
Ref.~\cite{PhysRevLett.105.223201}.

Beginning with the 2D limit ($\epsilon_b/\omega_z =0$), we determine
the energy of the tetramer compared to the trimer and dimer energies
(see Fig.~\ref{fig:undressed}).  Following the transition from a dimer
to a trimer at mass ratio $m_\up/m_\down\simeq3.33$, we find a
trimer-tetramer transition at $m_\up/m_\down\simeq5.0$.  In principle,
we can use Eq.~\eqref{eq:np1} to consider bound states of even larger
$N$, but the problem quickly becomes intractable numerically for
$N>3$.  However, we conjecture that composite bound states of larger
$N$ become possible as $m_\up/m_\down$ is increased, since the
relative importance of the centrifugal barrier between heavy particles
(which goes as $1/m_\up$) diminishes compared with the effective
attractive potential induced by the light particle ($\sim1/m_\down$).

Perturbing away from the 2D limit, we find that the trimer-tetramer
transition shifts to larger $m_\up/m_\down$ with increasing
$\epsilon_b/\omega_z$, as shown in Fig.~\ref{fig:phases}.  Eventually,
we expect to encounter the four-body Efimov effect in 3D for
$m_\up/m_\down > 13.4$~\cite{PhysRevLett.105.223201}. However, it
remains an open question whether our quasi-2D tetramers exist in 3D
below the critical mass ratio for Efimov physics.

To conclude, we have provided the first example of a universal,
non-Efimov tetramer involving three identical fermions.  Since this
quasi-2D tetramer exists for mass ratios $m_\up/m_\down$ as low as 5,
it could potentially be probed with ultracold $^6$Li-$^{40}$K
mixtures. Its small binding energy (Fig.~\ref{fig:undressed}) suggests
that it could appear as a resonance in atom-trimer interactions. For
instance, in the collision of a cloud of atoms and a cloud of trimers
under strong quasi-2D confinement, we expect the resonance to be
observable as a highly asymmetric density profile of scattered
atoms. This is similar to the proposal of Ref.~\cite{Levinsen2011} for
detecting an atom-dimer resonance.  In addition, the presence of
trimers and tetramers has implications for the many-body phases in
quasi-2D, particularly for the highly polarized Fermi
gas~\cite{Levinsen2012,inprep}.

We emphasize that although we have focussed on the $N+1$ problem in
quasi-2D, the form of Eq.~\eqref{eq:np1} is completely general and may
be extended to other shapes of the confining potential and/or
different dimensionalities.  For instance, in quasi-1D one would use
the $T$-matrix derived in Refs.~\cite{olshanii1998,RevModPhys.80.885},
along with appropriately redefined harmonic oscillator and momentum
indices.  Furthermore, the problem may be studied close to narrow
Feshbach resonances, characterized by a large effective range, by
using an energy-dependent scattering length \cite{Petrov2004}.
Finally, our work suggests that a two-channel model may be used to
model strongly confined quasi-2D Fermi systems in general.

{\em Note added:} After the submission of this manuscript, a similar
non-Efimov tetramer was predicted to exist in a 3D
geometry~\cite{Blume2012}.

\acknowledgments We gratefully acknowledge fruitful discussions with
Stefan Baur, Andrea Fischer, Pietro Massignan, Vudtiwat Ngampruetikorn,
and Dmitry Petrov. MMP acknowledges support from the EPSRC under Grant
No.\ EP/H00369X/1.  JL acknowledges support from a Marie Curie Intra
European grant within the 7th European Community Framework Programme.

\bibliography{Ref2D,2DRefs,morerefs}

\begin{thebibliography}{10}

\bibitem{Leggett1980}
A.~J. Leggett,  in {\em Modern Trends in the Theory of Condensed Matter},
  edited by A. Pekalski and J. Przystawa (Springer-Verlag, Berlin, 1980), p.\
  14.

\bibitem{Comte}
C. Comte and P. Nozi\`eres, J.\ Physique {\bf 43},  1069  (1982).

\bibitem{nozieres1985}
P. Nozi\`eres and S. Schmitt-Rink, J. Low Temp. Phys. {\bf 59},  195  (1985).

\bibitem{sademelo1993}
C.~A.~R. {S\'a de Melo}, M. Randeria, and J.~R. Engelbrecht, Phys. Rev. Lett.
  {\bf 71},  3202  (1993).

\bibitem{trimer}
O.~I. Kartavtsev and A.~V. Malykh, J. Phys. B: At. Mol. Opt. Phys. {\bf 40},
  1429  (2007).

\bibitem{PhysRevLett.103.153202}
J. Levinsen, T.~G. Tiecke, J.~T.~M. Walraven, and D.~S. Petrov, Phys. Rev.
  Lett. {\bf 103},  153202  (2009).

\bibitem{Mathy:2011ys}
C.~J.~M. Mathy, M.~M. Parish, and D.~A. Huse, Phys. Rev. Lett {\bf 106},
  166404  (2011).

\bibitem{orso2010}
G. Orso, E. Burovski, and T. Jolicoeur, Phys. Rev. Lett. {\bf 104},  065301
  (2010).

\bibitem{Efimov}
V.~N. Efimov, Nucl.~Phys.~A {\bf 210},  157  (1973).

\bibitem{Hammer2007}
H.-W. Hammer and L. Platter, Eur. Phys. J. A {\bf 32},  113  (2007).

\bibitem{stecher2009}
J. {von Stecher}, J.~P. D'Incao, and C.~H. Greene, Nature Phys. {\bf 5},  417
  (2009).

\bibitem{Ferlaino2008}
F. Ferlaino {\it et~al.}, Phys. Rev. Lett. {\bf 101},  023201  (2008).

\bibitem{stecher2010}
J. von Stecher, J. Phys. B: At. Mol. Opt. Phys. {\bf 43},  101002  (2010).

\bibitem{Bruch1979}
L.~W. Bruch and J.~A. Tjon, Phys. Rev. A {\bf 19},  425  (1979).

\bibitem{Platter2004}
L. Platter, H.-W. Hammer, and U.-G. Mei{\ss}ner, Few-Body Syst. {\bf 35},  169
  (2004).

\bibitem{LL}
L.~D. Landau and E.~M. Lifshitz, {\em Quantum Mechanics}
  (Butterworth-Heinemann, Oxford, UK, 1981).

\bibitem{PhysRevLett.105.223201}
Y. Castin, C. Mora, and L. Pricoupenko, Phys. Rev. Lett. {\bf 105},  223201
  (2010).

\bibitem{2DFermi_expt}
K. Martiyanov, V. Makhalov, and A. Turlapov, Phys. Rev. Lett. {\bf 105},
  030404  (2010).

\bibitem{PhysRevLett.106.105301}
B. Fr\"ohlich {\it et~al.}, Phys. Rev. Lett. {\bf 106},  105301  (2011).

\bibitem{sommer2011_2D}
A.~T. Sommer {\it et~al.}, Phys. Rev. Lett. {\bf 108},  045302  (2012).

\bibitem{Dyke2011}
P. Dyke {\it et~al.}, Phys. Rev. Lett. {\bf 106},  105304  (2011).

\bibitem{Zhang:2012uq}
Y. Zhang, W. Ong, I. Arakelyan, and J.~E. Thomas, Phys. Rev. Lett. {\bf 108},
  235302  (2012).

\bibitem{olshanii1998}
M. Olshanii, Phys. Rev. Lett. {\bf 81},  938  (1998).

\bibitem{nishida2010}
Y. Nishida and S. Tan, Phys. Rev. A {\bf 82},  062713  (2010).

\bibitem{2Dtrimer}
L. Pricoupenko and P. Pedri, Phys. Rev. A {\bf 82},  033625  (2010).

\bibitem{petrov2001}
D.~S. Petrov and G.~V. Shlyapnikov, Phys. Rev. A {\bf 64},  012706  (2001).

\bibitem{RevModPhys.80.885}
I. Bloch, J. Dalibard, and W. Zwerger, Rev. Mod. Phys. {\bf 80},  885  (2008).

\bibitem{Smirnov1962}
Y.~F. Smirnov, Nucl. Phys. {\bf 39},  346  (1962).

\bibitem{Skorniakov1956}
G.~V. Skorniakov and K.~A. Ter-Martirosian, Zh. Eksp. Teor. Phys. {\bf 31},
  775  (1956), [Sov. Phys. JETP {\bf 4}, 648 (1957)].

\bibitem{Minlos}
R. Minlos,  in {\em Proceedings of the Workshop on Singular Schr\"odinger
  Operators, Trieste, 29 September-1 October 1994}, edited by G. Dell'Antonio,
  R. Figari, and A. Teta (SISSA, Trieste, 1995).

\bibitem{Timmermanns1999}
E. Timmermans, P. Tommasini, M. Hussein, and A. Kerman, Physics Reports {\bf
  315},  199  (1999).

\bibitem{brodsky2006}
I.~V. Brodsky {\it et~al.}, Phys. Rev. A {\bf 73},  032724  (2006).

\bibitem{pricoupenko2011}
L. Pricoupenko, Phys. Rev. A {\bf 83},  062711  (2011).

\bibitem{Petrov2012}
D.~S. Petrov, V. Lebedev, and J.~T.~M. Walraven, Phys. Rev. A {\bf 85},  062711
   (2012).

\bibitem{mora2005}
C. Mora, A. Komnik, R. Egger, and A.~O. Gogolin, Phys. Rev. Lett. {\bf 95},
  080403  (2005).

\bibitem{Levinsen2011}
J. Levinsen and D.~S. Petrov, Eur. Phys. J. D {\bf 65},  67  (2011).

\bibitem{Levinsen2012}
J. Levinsen and S.~K. Baur, Phys. Rev. A {\bf 86},  041602  (2012).

\bibitem{inprep}
M.~M. Parish and J. Levinsen, in preparation.

\bibitem{Petrov2004}
D.~S. Petrov, Phys. Rev. Lett. {\bf 93},  143201  (2004).

\bibitem{Blume2012}
D. Blume, Phys. Rev. Lett. {\bf 109},  230404  (2012).

\end{thebibliography}

\end{document}